\documentclass{aastex63}

\received{2021 May 24}
\revised{2021 August 13}
\accepted{2021 August 17}
\published{2021 September 30}
\submitjournal{AJ}

\shortauthors{Ye et al.}

\begin{document}

\title{Extended Tidal Tails of IC 4756 detected by {\it Gaia} EDR3}

\correspondingauthor{Jingkun Zhao}
\email{zjk@nao.cas.cn}

\author{Xianhao Ye}
\affiliation{CAS Key Laboratory of Optical Astronomy, National Astronomical Observatories, Chinese Academy of Sciences, Beijing 100101, China}
\affiliation{School of Astronomy and Space Science, University of Chinese Academy of Sciences, Beijing 100049, China}

\author[0000-0003-2868-8276]{Jinkun Zhao}
\affiliation{CAS Key Laboratory of Optical Astronomy, National Astronomical Observatories, Chinese Academy of Sciences, Beijing 100101, China}

\author[0000-0003-1352-7226]{Jiajun Zhang}
\affiliation{CAS Key Laboratory of Optical Astronomy, National Astronomical Observatories, Chinese Academy of Sciences, Beijing 100101, China}
\affiliation{School of Astronomy and Space Science, University of Chinese Academy of Sciences, Beijing 100049, China}

\author[0000-0001-7609-1947]{Yong Yang}
\affiliation{CAS Key Laboratory of Optical Astronomy, National Astronomical Observatories, Chinese Academy of Sciences, Beijing 100101, China}
\affiliation{School of Astronomy and Space Science, University of Chinese Academy of Sciences, Beijing 100049, China}

\author{Gang Zhao}
\affiliation{CAS Key Laboratory of Optical Astronomy, National Astronomical Observatories, Chinese Academy of Sciences, Beijing 100101, China}
\affiliation{School of Astronomy and Space Science, University of Chinese Academy of Sciences, Beijing 100049, China}

\begin{abstract}

We report the discovery of emerged tidal tails around open cluster IC 4756 ($\sim$ 1 Gyr) based on 644 members identified from {\it Gaia} EDR3. Three-dimensional spatial positions, two-dimensional tangential velocities $\left( x, y, z, \kappa \cdot \mu_{\alpha}^{*}/\varpi, \kappa \cdot \mu_{\delta}/\varpi \right)$ are utilized to determine the co-moving member candidates of IC 4756. Using a Bayesian method, we correct the distance for each cluster member. Two tidal tails extend up to 180 pc and display a S-shape in $X^{\prime}Y^{\prime}$ space (Cartesian coordinates focused on cluster center). A clean sequence of our members in Color-Absolute-Magnitude Diagram (CAMD) indicates the coeval population and matches perfectly with the PARSEC isochrone with age from Bossini et al. (2019). Mass segregation is detected in this cluster as well. Finally, we derive the tidal radius and core radius of IC 4756 about $12.13$ pc and $4.33 \pm 0.75$ pc, respectively.

\end{abstract}


\section{Introduction} \label{sec:intro}

Born in giant molecular clouds (GMCs) \citep{Lada03}, and probably being the origins of most stars (not all stars; \citealt{Ward18,Ward20}), clusters are frequently chosen as the tracers to obtain the knowledge of the dynamical evolution of the galaxies. Investigations about the evolution and interaction about open clusters (OCs) present a deeper insight of the origin and evolution of the Galactic stars. Even for the clusters emerged from same GMC, they still may enter separated dynamical paths \citep{Pang20}. Therefore, each OC can be a unique object to study. 

One interesting feature for OCs is the long tidal tails. For young OCs ($\lesssim$ 100 Myr), the relic structures of the GMCs they formed or the fast gas expulsion can be the explanations for such structures \citep{Meingast5,Pang21}. For relatively elder clusters ($\gtrsim$ 100 Myr), two-body relaxation or external interactions, such as disk shocks, may be the main causes of the stripped members in tidal tails \citep{Carrera19}. \cite{DinnbierKroupa20} classified two kinds of tidal tails, tail I : formed because of gas expulsion; tail II : resulted from evaporation. A number of young OCs have been detected with tail-like structures, including IC 2391, IC 2602, NGC 2451A, NGC 2547 \citep{Meingast5} and NGC 2232, NGC 2451B \citep{Pang21} (NGC 2547 has been analyzed in both \citealt{Meingast5} and \citealt{Pang21}). In addition, more and more relatively elder OCs have been revealed with tidal tails, such as Hyades \citep{Meingast1,Roser1,Jerabkova21}; NGC 6774 \citep{Yeh19,Pang21}; Coma Berenices \citep{Furnkranz19,Tang19}; Praesepe \citep{Roser2}; NGC 2682 \citep{Carrera19,Gao20a}; Blanco 1 \citep{Zhang20,Meingast5,Pang21}; newly discovered OC UBC 274 \citep{Castro-Ginard20}; NGC 2506 \citep{Gao20b}; $\alpha$ Per \citep{Nikiforova20,Meingast5}; Platais 9, Messier 39 \citep{Meingast5}; NGC 6633 \citep{Pang21}; NGC 2516 \citep{Meingast5,Pang21}; NGC 752 \citep{Bhattacharya21}. Simulations \citep{Chumak&Rastorguev06,Just09,Kupper10,Ernst11,Jerabkova21} steadily promoted our awareness of the properties of tidal tails. However, due to the low number density comparing to the globular cluster, and the varying velocity components between cluster center and tails \citep{Jerabkova21}, it is not easy to identify the stripped tails members for OC, especially obscured by the accuracy of parallax and absence of radial velocity. 

IC 4756 is a relative old OC with age 890 $\pm$ 70 Myrs \citep{Strassmeier15} or $\log \mathrm{t} = 8.987$ \citep{Bossini19} covered by non-homogeneous dust \citep{Strassmeier15}. Its metallicity is close to solar value, [Fe/H] = -0.02 $\pm$ 0.01 \citep{Ting12}, [Fe/H] = -0.01 $\pm$ 0.10 \citep{Bagdonas18}. Its distance is about 478 pc \citep{Cantat-Gaudin18} to the Sun. Compared to the OCs with detected tidal tails : 1) IC 4756 is younger than NGC 752 ($\sim$ 1.34 Gyr; \citealt{Agueros18}) NGC 2506 ($\sim$ 2 Gyr; \citealt{Rangwal19}), NGC 6774 ($\sim$ 2.5 Gyr; \citealt{Curtis13}), UBC 274 ($\sim$ 3 Gyr; \citealt{Castro-Ginard20}) and NGC 2682 ($\sim$ 3.6 Gyr; \citealt{Bossini19}), but elder than the others; 2) farther than most of them (except NGC 2682, NGC 2506 and UBC 274. Considering the age and the low Galactic latitude $b = 5.325^{\circ}$ \citep{Cantat-Gaudin18} of this cluster, the tidal tails (tail II) are most likely to be formed already.

The aim of this paper is to detect tidal tails around IC 4756. Meanwhile, its radial density profile and the analysis of mass segregation are also studied. The structure of this paper is built as follows : Sec. \ref{sec:data} illustrates the constraints to extract initial data sample from {\it Gaia} EDR3; the methods of members selection, correction for distance, and the tails detection are described in Sec. \ref{methods}; In Sec. \ref{results}, we present the photometric properties of the member candidates, the diagnosis of mass segregation and radial density profile. Finally, our main conclusions are summarized in Sec. \ref{summary}.

\section{Data} \label{sec:data}

With the known information of IC 4756 \citep{Cantat-Gaudin18}, the initial sample can be restricted with the position on sky and parallax $\left( \alpha, \delta, \varpi \right)$ accordingly. To serve our purpose of finding co-moving tidal structures, restrictions of the relative errors of proper motion $\left( \mu_{\alpha}^{*}, \mu_{\delta} \right)$ are up to $30\%$. However, considering the small value of proper motion in the right ascension for IC 4756  ($\mu_{\alpha}^{*} \sim 1.26$ $\mathrm{mas} \cdot \mathrm{yr} ^{-1}$), limits on relative errors may lead to absence of stars that have small accurate $\mu_{\alpha}^{*}$ and potentially are the member candidates. Therefore, we retain the stars which $\sigma_{\mu_{\alpha}^{*}} < 0.378$ $\mathrm{mas} \cdot \mathrm{yr} ^{-1}$. This limit is the $30\%$ of the $\mu_{\alpha}^{*}$ of IC 4756 and we can retain stars $\mu_{\alpha}^{*} \sim 0$. The relative errors of parallax are down to $10\%$ to ensure the quality of distances derived from parallax. According to the range of $\varpi$, the minimum error cut on $\varpi$ is about 0.143 mas, which should not bring the same issue as in $\mu_{\alpha}^{*}$. In addition, the renormalized unit weight error (RUWE; \citealt{Lindegren18tech}), provided along with {\it Gaia} EDR3 \citep{Gaia_summary,Lindegren20}, are used to get the sources with fine astrometric solutions. The initial sample of about half a million stars near IC 4756 is extracted from EDR3 based on the criteria as following List \ref{cut0}:
\begin{itemize}\label{cut0}
  \item 270 $<$ $\alpha$ $<$ 295, -10 $<$ $\delta$ $<$ 20, 1/0.700 $<$ $\varpi$ $<$ 1/0.300;
  \item \texttt{parallax\underline{ }over\underline{ }error} $>$ 10;
  \item $\sigma_{\mu_{\alpha}^{*}}/|\mu_{\alpha}^{*}| < 0.3$ or $\sigma_{\mu_{\alpha}^{*}} < 0.378$;
  \item $\sigma_{\mu_{\delta}}/|\mu_{\delta}| < 0.3$;
  \item RUWE $<$ 1.4 .
\end{itemize}
496,103 sources contain three {\it Gaia} photometric passbands, and error cuts for $\left( \varpi,\mu_{\alpha}^{*},\mu_{\delta} \right)$ may lead to significant incompleteness for those $G \ge 18$ mag in this sample. However, only about $6\%$ stars with radial velocity from {\it Gaia} DR2. Therefore, 5D phase space with position and tangential velocity is adopted to extract co-moving stars associated with IC 4756. We use distance from \cite{Bailer-Jones21} in the following calculation of spatial coordinates, but for those without this distance, $1/\varpi$ is adopted. Parallax zero point is calculated using the code from \cite{Lindegren20-zp}.

\section{Methods} \label{methods}

\subsection{Member Selections} \label{members}

\begin{figure*}[t!]
	\includegraphics[width=1.00\textwidth]{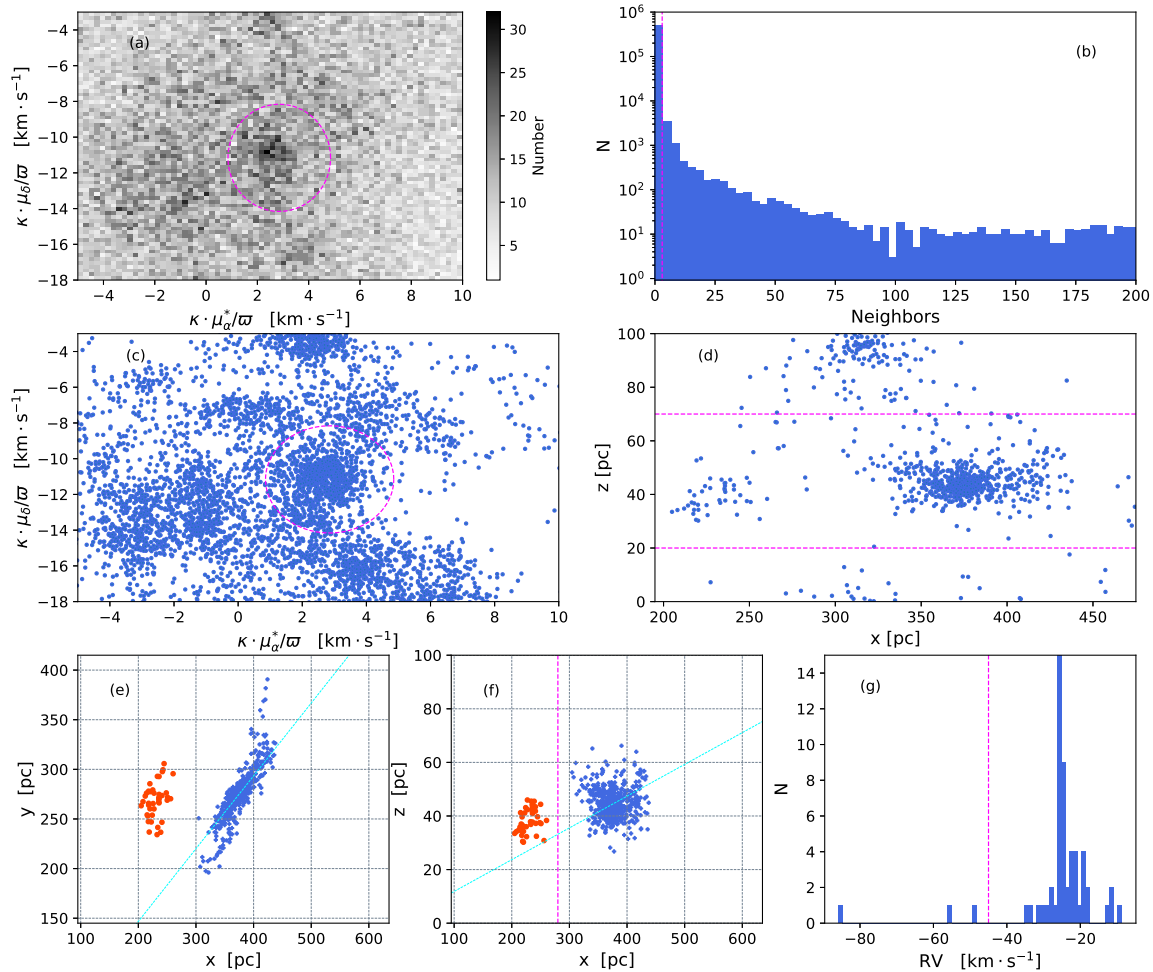}
	\caption{Steps to obtain cluster memberships for IC 4756. (a) Density map around IC 4756, a distinct over density is revealed inside the dashed magenta ellipse, which is centered at the predicted center calculated with parameter from \cite{Cantat-Gaudin18}. (b) The histogram plot of so called 5D neighbors for all 496,103 sources from {\it Gaia}. A number of individuals have neighbors $\le 3$, indicated by dashed magenta line. (c) Tangential velocities for stars that neighbors $> 3$. A bulk of stars that within the ellipse are into the next steps. (d) Overdensity in Galactic Cartesian coordinates $xz$ plane. The center of this right-handed coordinates system is the Sun, positive $x$ pointing to the Galactic center, positive $y$ the Galactic rotation directions, and $z$ to north Galactic pole. $20$ pc $< z < 70$ pc cut leaves a more clean members. (e) Two groups clustered by \texttt{HDBSCAN} are shown in different colors in $x-y$ plane, IC 4756 corresponding to the blue plus signs. Line-of-sight is indicated by the cyan line. (f) Two groups from (e) in $x-z$ plane. Dashed magenta line separates them. (g) Distribution of {\it Gaia} DR2 radial velocities and sources on the left side of the dashed magenta line are removed. \label{fig:extraction}}
\end{figure*}

\begin{figure*}[t!]
	\includegraphics[width=0.50\textwidth]{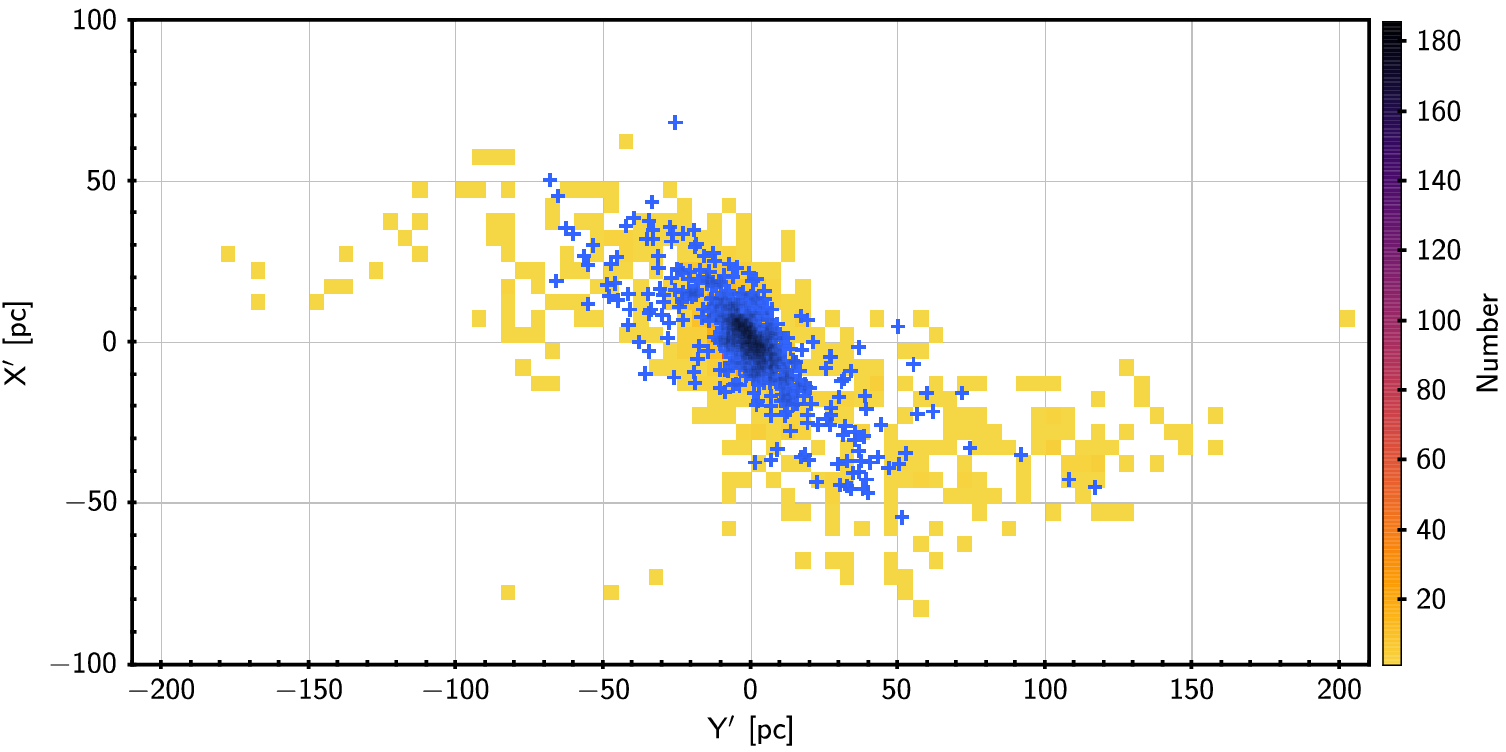}
  \includegraphics[width=0.50\textwidth]{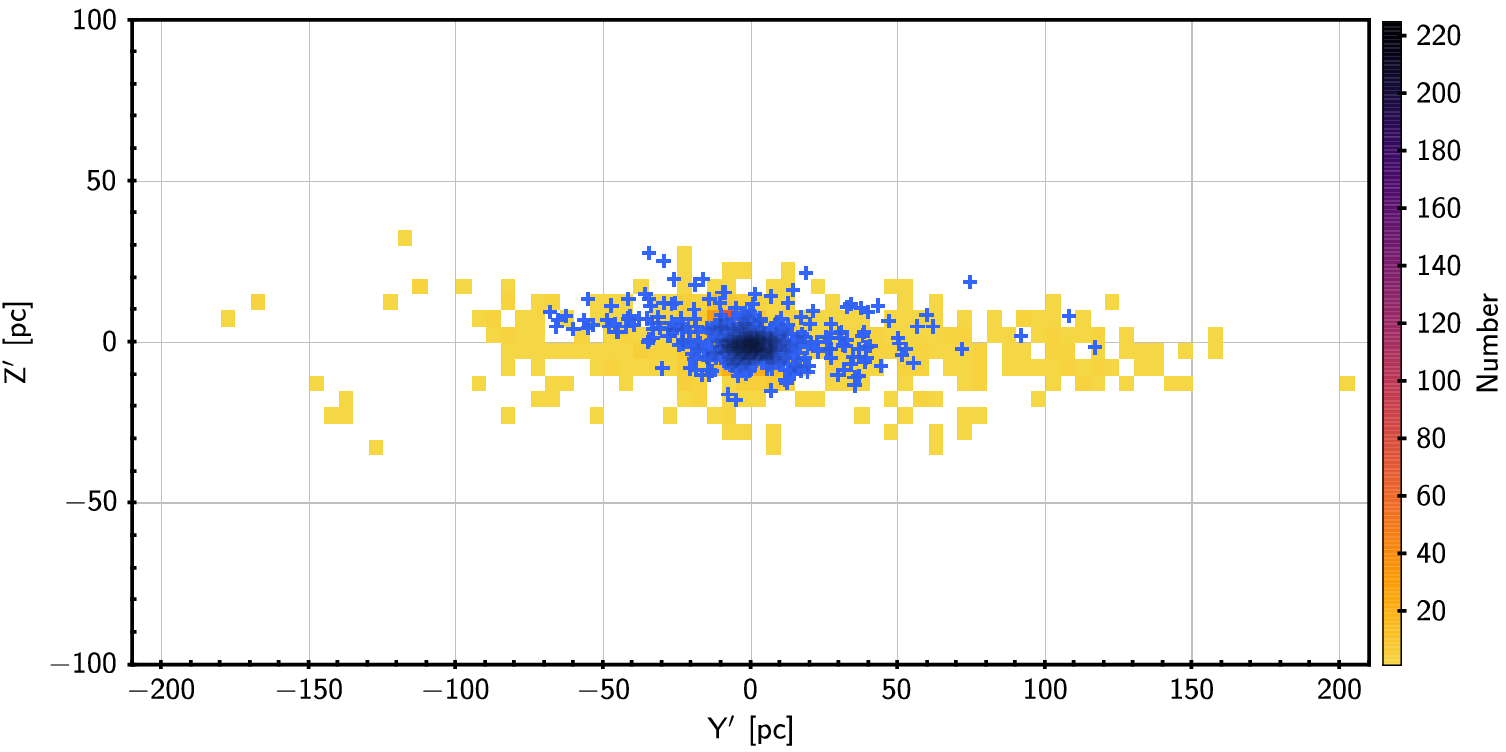}
	\caption{Member candidates for IC 4756 compared with already discovered tidal tails in $\left( Y^{\prime}, X^{\prime} \right)$ and $\left( Y^{\prime}, Z^{\prime} \right)$. The color bars show the number density of tidal tails from literatures. Blue plus signs are IC 4756 memberships. \label{fig:comparison}}
\end{figure*}

Panel (a) of Fig. \ref{fig:extraction} presents the tangential velocities distribution of the 496,103 sources around the center of IC 4756 in tangential velocity space. Dashed magenta ellipse fences out the obvious over-density, indicating the potential cluster co-moving members. It is hard to distinguish the co-moving tidal tails only with spatial coordinates due to the low number density. Therefore, we first define 5D neighbors to roughly remove a large amount of stars that do not belong to any over-density in phase space. We calculate neighbors for all 496,103 sources, which is defined in phase space $\left( x, y, z, \kappa \cdot \mu_{\alpha}^{*}/\varpi, \kappa \cdot \mu_{\delta}/\varpi \right)$, same as Eq. (1) in \cite{Ye21} and the reference therein. In short, for each source, its neighbors is defined within a radius of $15$ pc in $\left( x, y, z \right)$ and a radius of $1.1$ $\mathrm{km}$ $\mathrm{s}^{-1}$ in $\left( \kappa \cdot \mu_{\alpha}^{*}/\varpi, \kappa \cdot \mu_{\delta}/\varpi \right)$. Panel (b) of Fig. \ref{fig:extraction} shows the histogram of neighbors. The dashed magenta line represent neighbors $=3$. It is clear that numerous stars have neighbors less than 3. Thus we can roughly remove field stars and keep 7,312 stars with neighbors $> 3$ into the next procedure. Taking $\mu_{\alpha}^{*}, \mu_{\delta}$ and $\varpi$ from \cite{Cantat-Gaudin18} to estimate the center of tangential velocity $\left( \kappa \cdot \mu_{\alpha}^{*}/\varpi, \kappa \cdot \mu_{\delta}/\varpi \right)$ for IC 4756, we then extract stars within the dashed magenta ellipse in panel (c), which cover the full overdensity in tangential velocity space and only contain a small fraction of contamination. The ellipse is described by the second item in List \ref{cut1}, and it is the same as in Fig .\ref{fig:extraction} panel (a). An evident overdensity is unfolded in Galactic Cartesian coordinates (centered at Sun) $xz$ plane in panel (d), corresponding to our target cluster. Positive $x$ in this frame points toward the Galactic center, $y$ directs the local rotation, and $z$ points to the Galactic north pole. Considering the thin distributions in $z$ direction, we cut $20$ pc $< z < 70$ pc (between dashed magenta lines) to avoid a number of stars that have no significant differences in tangential velocities but scattered in spatial coordinates. After the above steps, 725 stars left. Then \texttt{HDBSCAN} \citep{Campello13,McInnes17} with \texttt{min\underline{ }cluster\underline{ }size} $=20$ is used to obtain the members of IC 4756 in $\left( x,y,z \right)$ space. This algorithm detected two clusters : one is exactly IC 4756 (displayed with blue plus signs in panel (e-f) of Fig. \ref{fig:extraction})  and the other one corresponds to the clump shown in panel (d) around $\left( 230,40 \right)$ in $xz$ plane (marked as orange dots in panel (e) and (f)). The latter is beyond the extent of IC 4756 and its tidal tails, so we do not focus on it in this paper. The cyan lines indicate the line of sight in Galactic coordinates, and the magenta line in panel (f) separates the two \texttt{HDBSCAN} identified clusters. The last panel shows the {\it Gaia} DR2 radial velocity distribution. Most stars locates around $-24.74$ $\mathrm{km} \cdot \mathrm{s}^{-1}$ \citep{Soubiran18}. Three sources away from the distribution center are discarded (on the left side of the dashed line). At last, we obtain 644 cluster member stars. The corresponding cuts described above are listed in List \ref{cut1} below.

\begin{itemize}\label{cut1}
	\item neighbors $>$ 3 $\longrightarrow$ 7,312 stars remaining;
	\item $\frac{\left( \kappa \cdot \mu_{\alpha}^{*}/\varpi - 2.85 \right)^2}{2^2} + \frac{\left( \kappa \cdot \mu_{\delta}/\varpi + 11.16 \right)^2}{3^2} \le 1$ $\longrightarrow$ 1,019 stars remaining;
	\item $20$ pc $< z < 70$ pc $\longrightarrow$ 725 stars remaining;
  \item \texttt{HDBSCAN}; 
  \item cut on radial velocity $\longrightarrow$ 644 stars remaining\,.
\end{itemize}

\subsection{Correction for Distance}
The elongated morphology of our extracted members are influenced a lot by accuracy of individual distance as indicated by the line-of-sight in panel (e) of Fig. \ref{fig:extraction}. Even we have limited the \texttt{parallax\underline{ }over\underline{ }error} to make a 10\% relative error cut in $\varpi$, and adapt the distance from \cite{Bailer-Jones21} to replace the simply inverted $\varpi$, the observation effect is still not completely eliminated. To moderate this impact, we follow the Bayesian approach detailed in \cite{Bailer-Jones15} and introduce a prior used in \cite{Carrera19} to calculate the likelihood. This prior can be described by a sum of two parts, indicating cluster members and field stars respectively. In Eq. \ref{prior}, the former exponentially decreasing term represents the field stars and the latter one indicates the cluster members.
\begin{eqnarray}\label{prior}
  P\left( d \right) \propto C \cdot d^{2} e^{-\frac{d}{8 \mathrm{[kpc]}}} + \left( 1-C \right) \cdot \frac{1}{\sqrt{2\pi \sigma_{d}^{2}}}e^{-\frac{(d-d_{0})^2}{2 \sigma_{d}^{2}}}
\end{eqnarray}
$d_{0}$ is the predicted mean distance of cluster members, adopted as the inverted parallax from \cite{Cantat-Gaudin18}. $\sigma_{d}$ is the deviation of the distances between member stars and its cluster center, as used in \cite{Pang20,Pang21}. Differed from \cite{Pang20,Pang21}, we set unique coefficient $C$ for each individual. The coefficients $C$ is the contamination fraction same as Eq. (1) in \cite{Meingast5}, like $\rho_{\mathrm{field}}/\left( \rho_{\mathrm{cluster},i}+\rho_{\mathrm{field}} \right)$. In a nutshell, the $\rho_{\mathrm{field}}, \rho_{\mathrm{cluster},i}$ are the volume number density of field stars and cluster member star $i$. For each cluster member, $\rho_{\mathrm{cluster},i}$ is calculated within a radius of the distance to its seventh ($7$th) nearest neighbor among 644 cluster candidates. Following \cite{Meingast5}, the sample of field sources is chosen as the inner stars of the ellipse in panel (a) of Fig. \ref{fig:extraction} with the final cluster member stars removed. Because those stars are co-moving with IC 4756, but not the member candidates. $\rho_{\mathrm{field}}$ is derived with same definition for $\rho_{\mathrm{cluster},i}$ but is set as the median value of this sample.

\subsection{Extended Tails For IC 4756}

\begin{figure*}[t!]
	\includegraphics[width=0.50\textwidth]{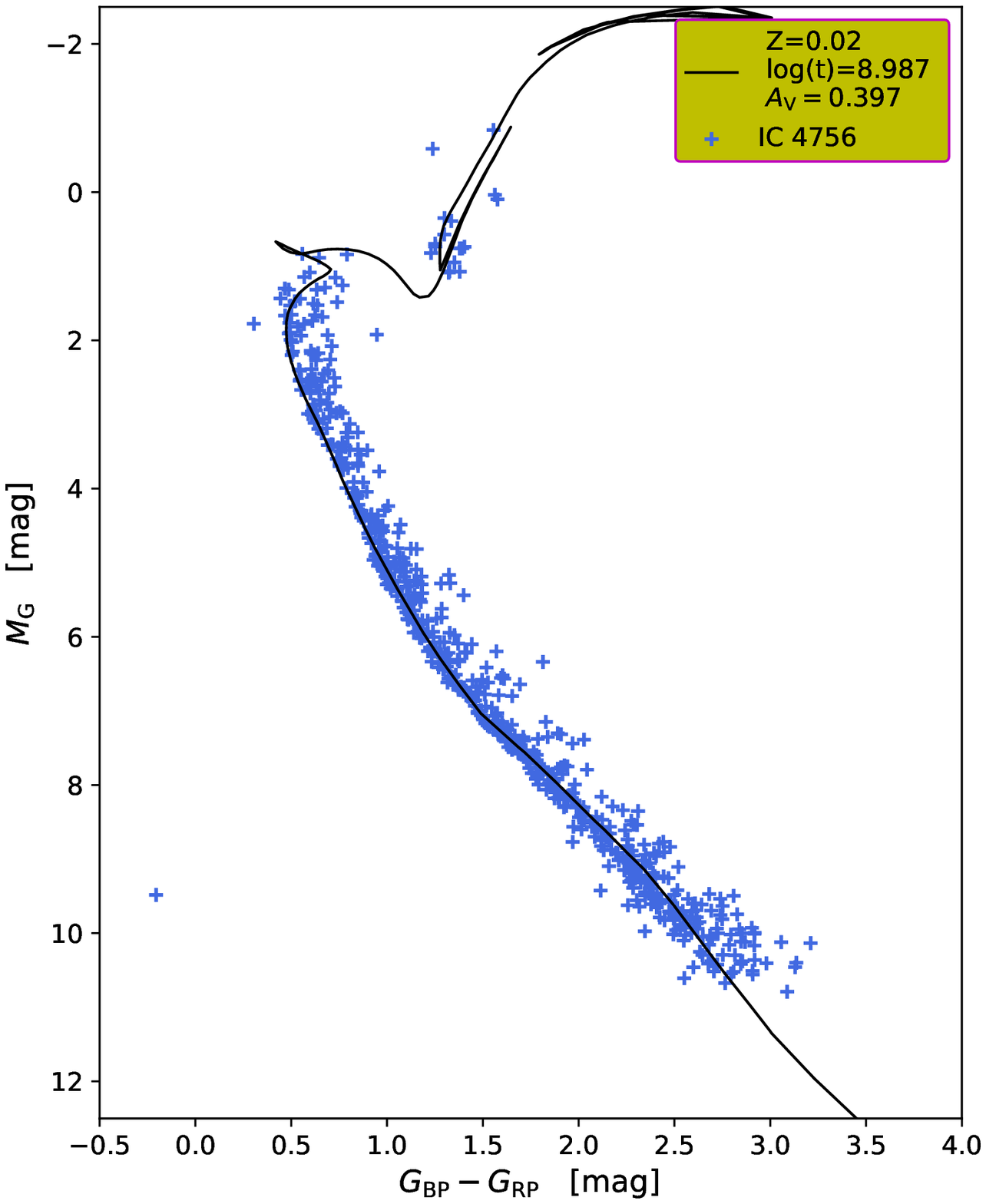}
  \includegraphics[width=0.50\textwidth]{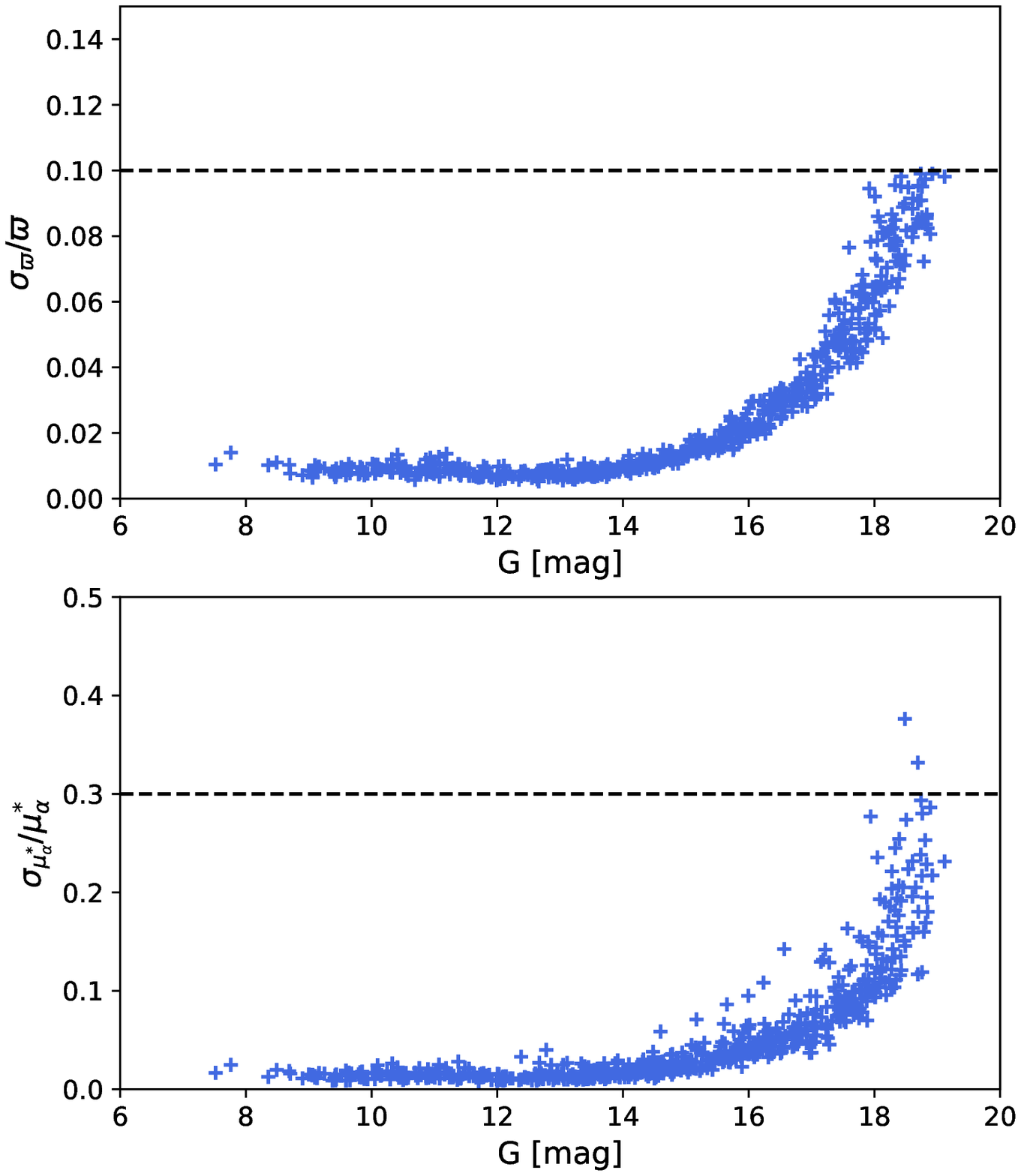}
	\caption{Left : CAMD of 644 members of IC 4756. The blue plus signs represent cluster members and  PARSEC isochrone is plotted as black line. Right : Magnitude $G$ vs. relative errors ($\varpi$ and $\mu_{\alpha}^{*}$) of those members. Black dashed lines are corresponding the $30\%$ cuts used in data extraction.  \label{fig:cmd}}
\end{figure*}

The revised distance for each individual has then been utilized in a left-handed Galactocentric Cartesian coordinates computations. Sun is set at height $Z_{\odot} = 0.027$ kpc \citep{Chen01} and at a radius of $R_{\odot} = 8.3$ kpc \citep{Gillessen09}. To better compare memberships spatial distribution with the tidal tails in literatures, we rotate and translate the Galactocentric Cartesian coordinates to focus on cluster center. This new coordinates $X^{\prime}$ points from Galactic center to cluster center, $Y^{\prime}$ directs the Galactic rotation in this local, and $Z^{\prime}$ still points to the north Galactic pole. In Fig. \ref{fig:comparison}, our member candidates are compared to the detected tidal tails of Hyades \citep{Meingast1} and Praesepe \citep{Roser2}, which are transformed in this same coordinates as well. Member candidates of IC 4756 present quite similar distribution as the tails in \cite{Meingast1} and \cite{Roser2} in $\left( X^{\prime}, Y^{\prime}, Z^{\prime} \right)$ space. The observation effect is effectively diminished, compared with panel (e) in Fig. \ref{fig:extraction}, and two tails stand out, apparently. But the extent of tails is smaller than \cite{Meingast1} and \cite{Roser2}, only up to 180 pc for total length. In addition, the leading tail is slightly longer than the trailing tail. All the member candidates demonstrate a S-shape in $\left( X^{\prime},Y^{\prime} \right)$. As shown in Fig. \ref{fig:comparison}, most stars are gathered around dozens of parsecs and a relatively much lower number density in the tails.

\section{Results} \label{results}

\subsection{Color-Absolute-Magnitude Diagram}

All 644 member candidates are drawn in color-absolute magnitude diagram (CAMD) as $G_{\mathrm{BP}} - G_{\mathrm{RP}}$ and $M_{\mathrm{G}}=G-5\lg\left( d/10 \right)$, where $d$ is the corrected distance in pc. PARSEC (version 1.2S) isochrone \footnote{\url{http://stev.oapd.inaf.it/cgi-bin/cmd}} \citep{Bressan12,Chen14,Chen19} with $Z$ = 0.02, $\log \mathrm{t} = 8.987$, $A_{\mathrm{V}} = 0.397$ \citep{Bossini19} is together shown in left panel of Fig. \ref{fig:cmd}. A clean CAMD is fitted excellently with the PARSEC isochrone for main sequence and giants branch. Noted that we do not apply any extinction for these {\it Gaia} EDR3 passbands, and $A_{\mathrm{V}} = 0.397$ is used in PARSEC isochrone, so the sequence of cluster members may be less scattered when accurate extinction parameters are adopted. Another proof for the correctly clean member selection is a two-dimensional KS test \citep{Peacock83}. It can inspect member candidates to verify that the members are not casually selected from the initial sample in Sec. \ref{sec:data}. Following \cite{Meingast1}, we get a mean $p-$value about $10^{-18}$ between our candidates and the randomly collections (all have same length as our candidates selection) taken from initial sample. The mean value is obtained from 10,000 tests. The mean value between two random sets is typically about 0.52. The right panel of Fig. \ref{fig:cmd} shows that our member candidates may suffer incompleteness for fainter stars $G \approx 18$ mag ($M_{\mathrm{G}} \approx 9.7$ mag), according to the relative errors cuts in Sec. \ref{sec:data}. 

\subsection{Mass Segregation}

\begin{figure}[htbp!]
  \centering
	\includegraphics[width=0.50\textwidth]{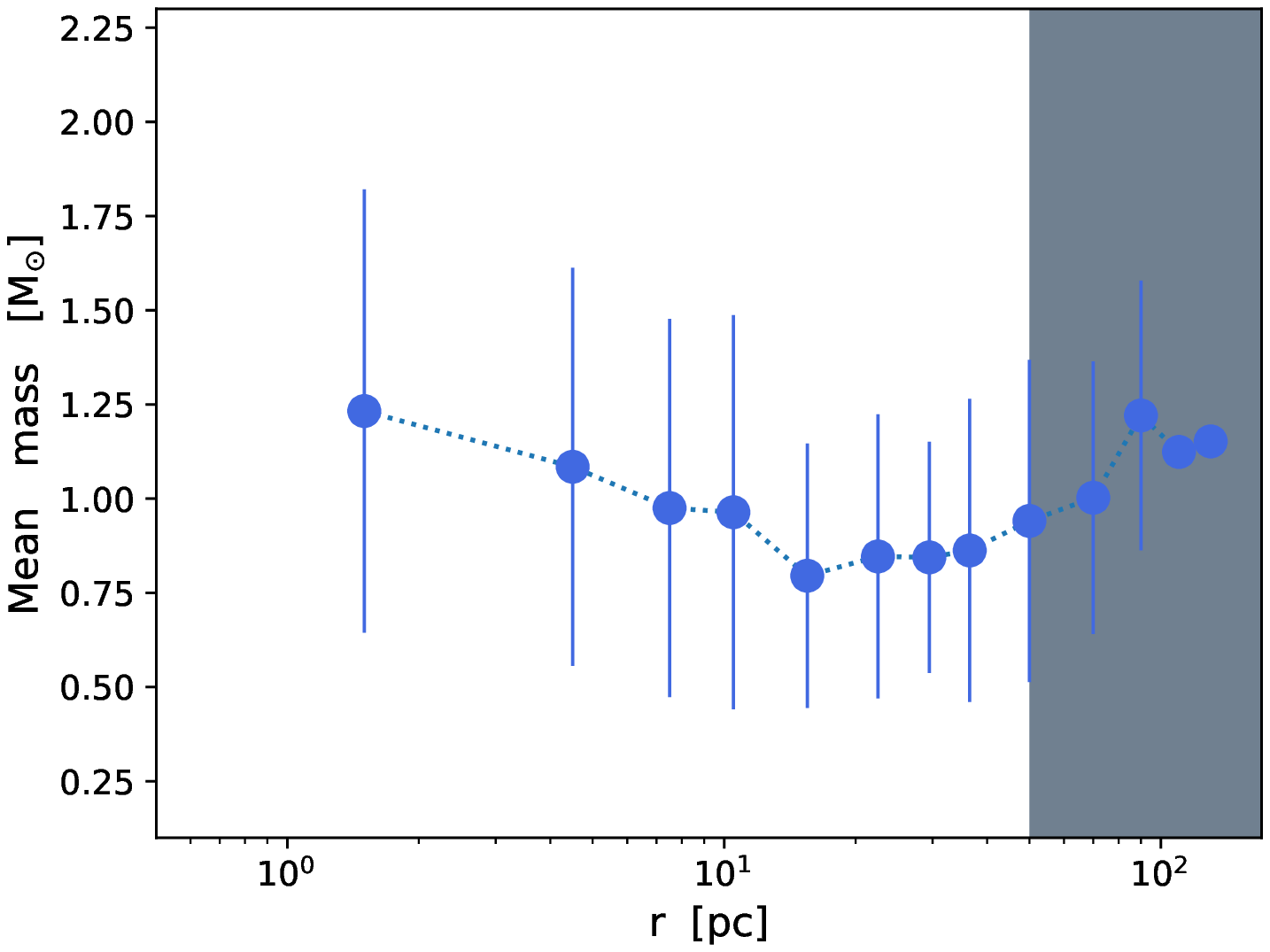}
  \includegraphics[width=0.50\textwidth]{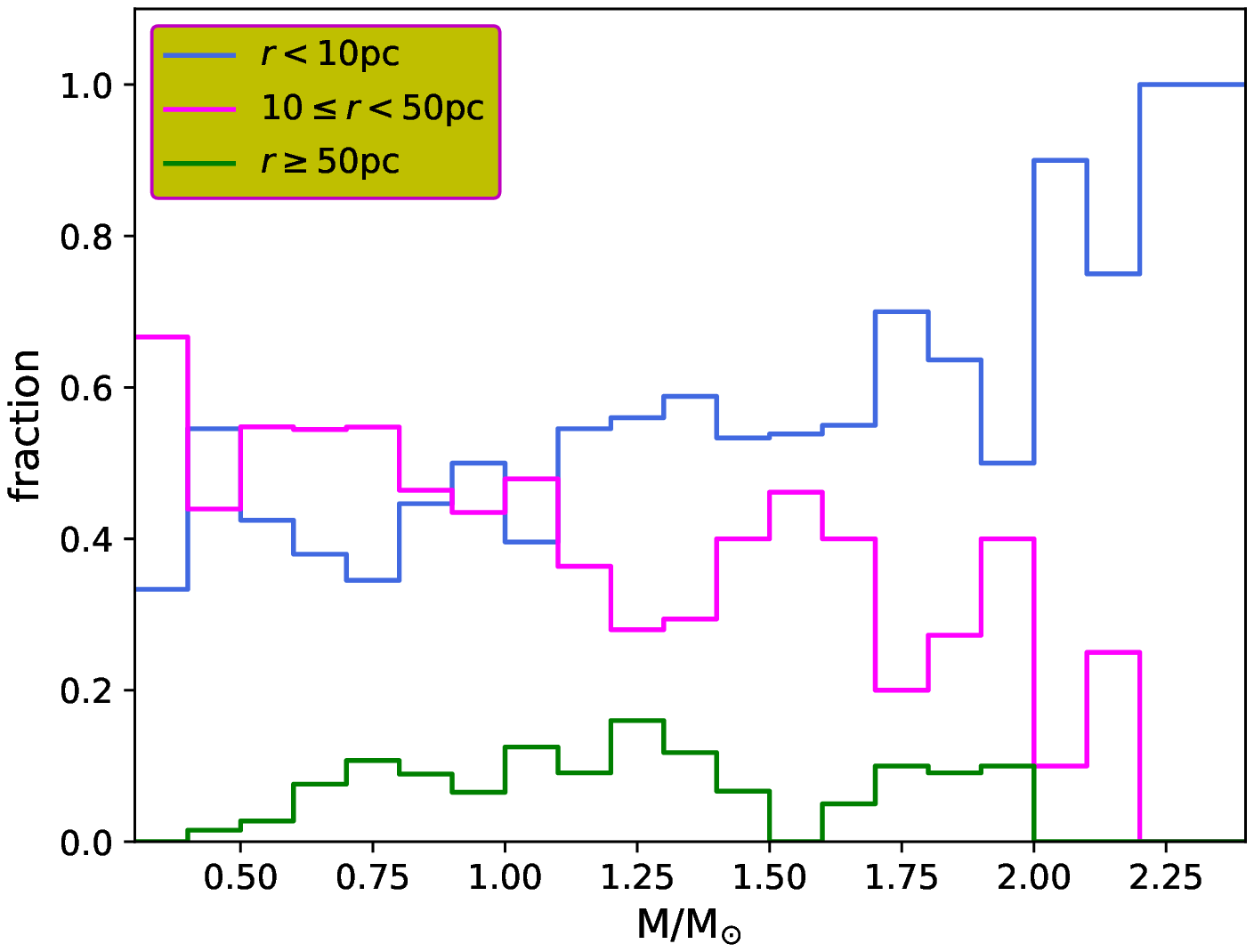}
	\caption{Top panel : Mean stellar mass of IC 4756 vs. distance from the cluster center $r$ in different annuli. The intervals are larger when stars are far away from the center. The standard deviation is presented as the error bar in each annulus. The grey shadow represents the area that $r \ge 50$ pc. Bottom panel : mass functions for $r < 10$ pc, $10 \le r < 50$ pc, and $r \ge 50$ pc, shown as blue, magenta and green lines, respectively. Y-axis is the fraction of numbers in each bin.  \label{fig:massseg}}
\end{figure}

Stars with different mass may be segregated in a cluster, where the more massive stars assembled in the core of the cluster and less massive individuals tends to evaporated from it \citep{HH98}. It might be from the primordial segregation \citep{Pavlik19} when they were formed, or dynamical consequence \citep{Allison09apjl}. \citealt{Allison09mnras} presented the approach without previous knowledge about the center position of the cluster to do a quantitative analysis for mass segregation. The mass segregation is detected by analyze the mean mass of each individual in different annuli. A rough estimate for stellar mass for each member of IC 4756 is derived through the CAMD and PARSEC isochrone. For each individual, its mass is adopted as the mean mass of two nearest stars from isochrone. The top panel of Fig. \ref{fig:massseg} presents the mean stellar mass as a function of distance from the cluster center. The distance of each annulus from the center is calculated with three-dimensional coordinates $\left( X^{\prime},Y^{\prime},Z^{\prime} \right)$ centered at $\left( 0,0,0 \right)$. From Fig. \ref{fig:massseg} top panel, we can see a slight tendency that the mean mass declines about dozens of pc away from the center. But due to the large dispersions shown as error bars in the figure, it is not convincing to conclude a mass segregation here. Besides, this tendency disappears out 50 pc, grey shadow. It might be the reason that few members are beyond 50 pc ($93 \%$ stars are within 50 pc). The bottom panel of Fig. \ref{fig:massseg} presents the three mass functions of the members for different distance from the center. The members are divided into three groups according to the distance from center $r$ : $r < 10$ pc, $10 \le r < 50$ pc, and $r \ge 50$ pc, which are presented as blue, magenta, and green lines, respectively. The $y$ axis is the fraction of numbers in each bin. Although it displays no inclination for the group of $r \ge 50$ pc, this figure certainly reveals the upward and downward trends of $r < 10$ pc and $10 \le r < 50$ pc MFs for more massive stars. They agree with the tendency shown in the top panel of Fig. \ref{fig:massseg}. Therefore, we conclude the mass segregation in IC 4756.

\subsection{Radial Density Profile} \label{rdp}

\begin{figure}[t!]
	\includegraphics[width=0.47\textwidth]{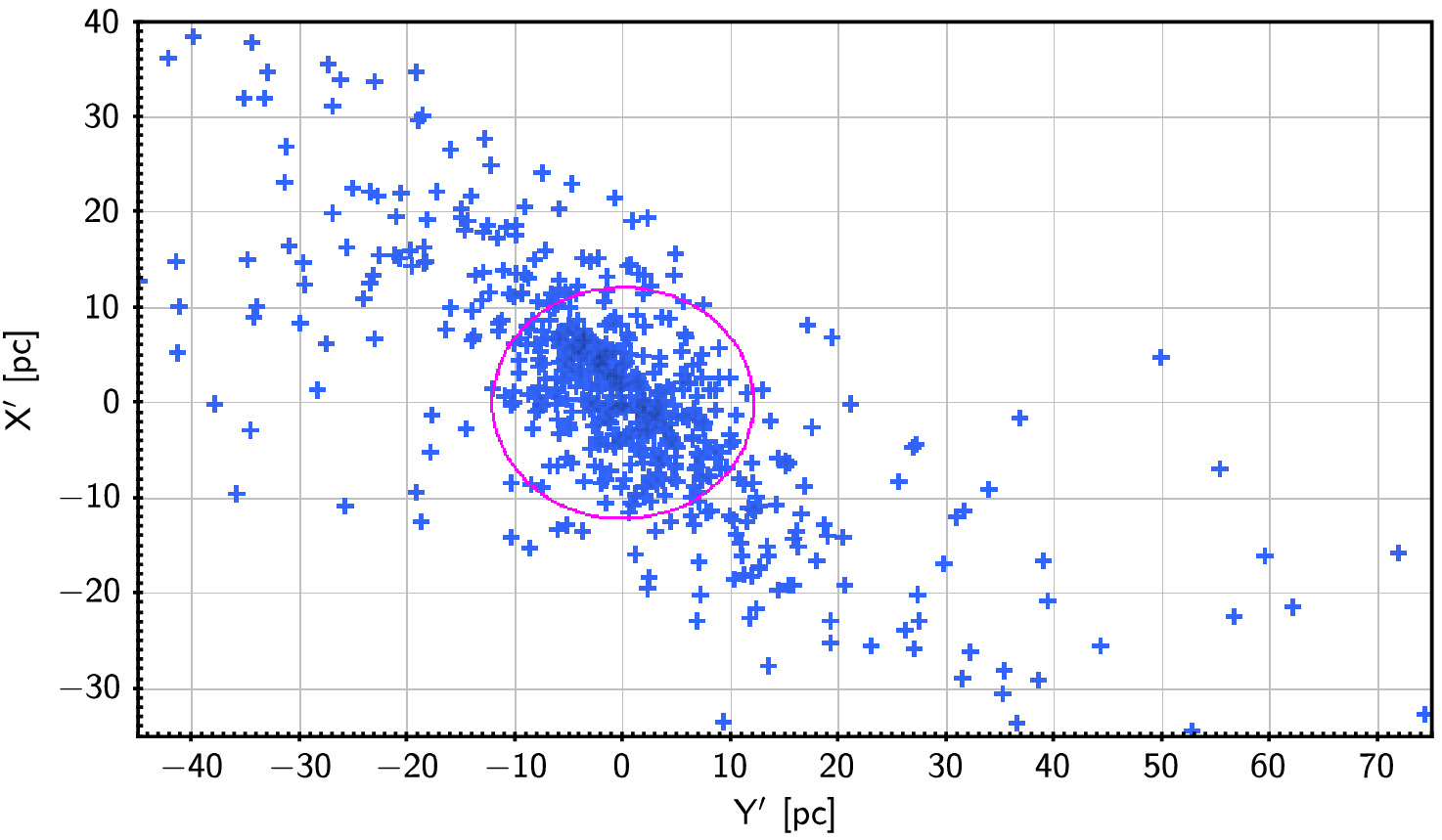}
  \includegraphics[width=0.50\textwidth]{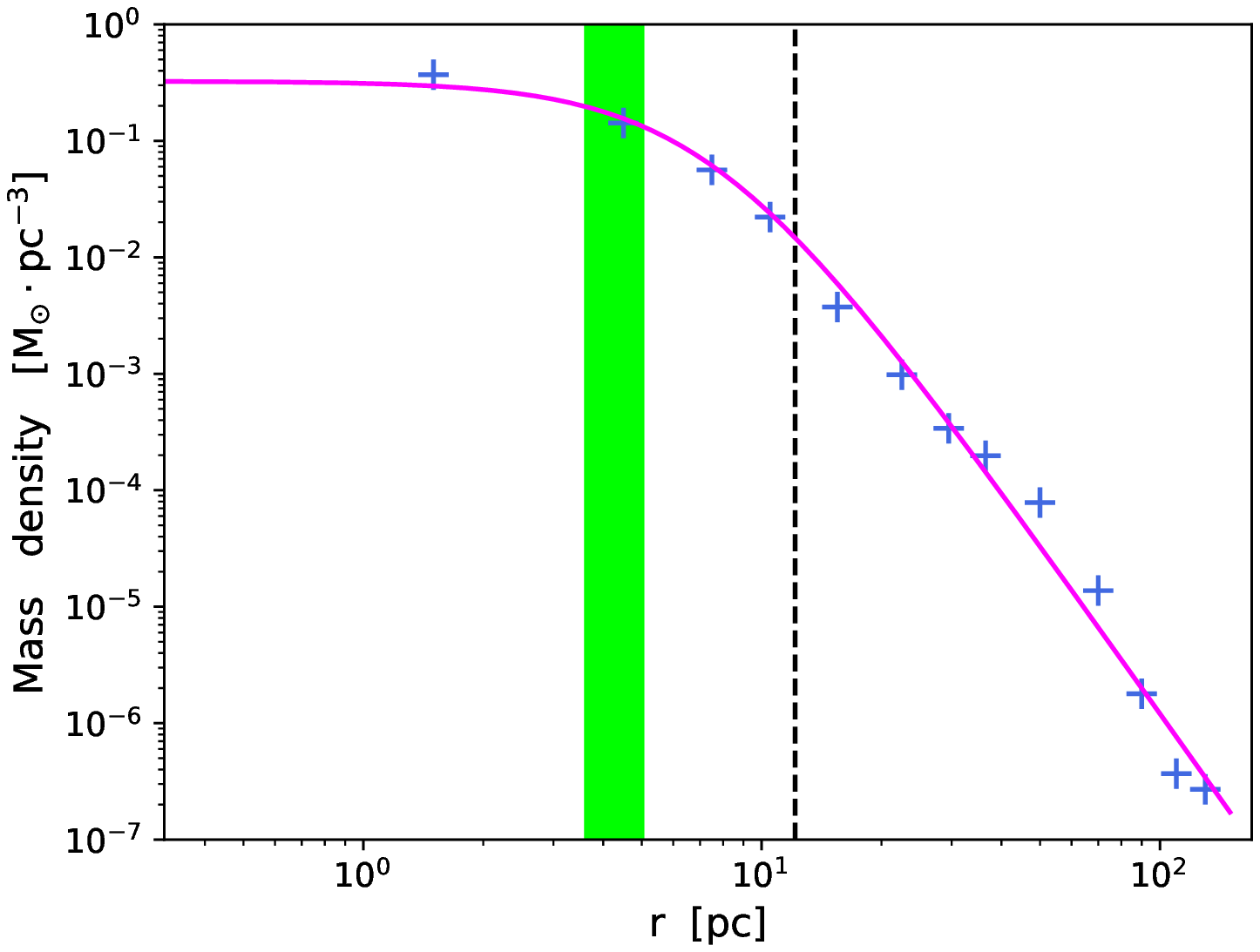}
	\caption{Top panel : The distribution of member stars of IC 4756 in $X^{\prime}Y^{\prime}$ space. Tidal radius is presented as the magenta circle.  Bottom panel : Radial mass density of IC 4756 vs. distance from the cluster center in different annuli. The intervals of distance are same as those in Fig. \ref{fig:massseg}. The magenta curve is the fitted model \citep{Elson87} with a core radius of $ 4.33 \pm 0.75$ pc, shown as the green shadow. Tidal radius $r_{\mathrm{t}}$ is also presented as vertical black dashed line.  \label{fig:rt_and_rdp}}
\end{figure}

\cite{Kharchenko13} derived the core radii and the tidal radii of OCs according to a King density model \citep{King62}. However, considering the long extended structure in IC 4756, it is not appropriate to construct its radial density profile using the power law from \cite{King62}. In most cases, not all of them, the parameter of density law, tidal radius, is close to the Jacobi radius (distance between a star cluster center and the Lagrange points), which indicates the boundary of gravitationally bound and unbound stars in a star cluster. For our cluster members, we use Eq. \ref{rt_Pinfield}  \citep{Pinfield98} to estimate the value of this gravitationally bound radius, which is adopted as tidal radius in this paper. 
\begin{eqnarray}
  r_{\mathrm{t}} &=& \left[ \frac{G M_{\mathrm{total}}}{2\left( A-B \right)^2} \right]^{1/3} \,. \label{rt_Pinfield}
\end{eqnarray}
The $M_{\mathrm{total}}$ in Eq. \ref{rt_Pinfield} is the total mass of cluster members. Gravitational constant $G = 4.3 \times 10^{-6}$ $\mathrm{kpc}$ $( \mathrm{km}$ $\mathrm{s}^{-1} )^{2}$ $\mathrm{M}_{\mathrm{\odot}}^{-1}$, and the Oort constants $A$ = 15.3 $\pm$ 0.4 $\mathrm{kpc}^{-1}$ $\mathrm{km}$ $\mathrm{s}^{-1}$, $B$ = -11.9 $\pm$ 0.4 $\mathrm{kpc}^{-1}$ $\mathrm{km}$ $\mathrm{s}^{-1}$ are from \citealt{Bovy17}. In this way, we obtain the tidal radius $r_{\mathrm{t}} = 12.13$ pc, shown as the magenta circle in top panel of Fig. \ref{fig:rt_and_rdp}. We present the members closer to the center part in $X^{\prime}Y^{\prime}$, comparing to the left panel of Fig. \ref{fig:comparison}, in order to give a clearer view around the circle of tidal radius. Stars inside the circle are distributed more compactly. $56\%$ members concentrate inside $r_{\mathrm{t}}$, which contribute $60\%$ mass of IC 4756. As considerable parts of members and mass of IC 4756 are beyond the tidal radius and IC 4756 is about $1$ Gyr old, we presume that IC 4756 is undergoing disruption.

The radial density profile is also computed. It\textsl{} decreases drastically as the distance to cluster center increases, but with the long tidal tails, it is more appropriate to be compared with the power law from Elson, Fall, Freeman model \cite{Elson87}, hereafter EFF model. We take radial mass density instead of the surface brightness in EFF model, and the power law is presented in Eq. \ref{eff},
\begin{eqnarray} \label{eff}
  \rho\left( r \right) = \rho_{0} \cdot \left( 1 + \left( r/a \right)^2  \right)^{-\gamma/2} \,,
\end{eqnarray}
where $\rho_{0}$ is the mass density of a cluster in its center, $a$ is related to the core radius and $\gamma$ is the index of slope. The core radius $r_{\mathrm{c}}$ indicates the sphere where the surface brightness of a cluster falls by half of its central surface brightness. In our situation, we adopt the relation $\rho \left( r_{\mathrm{c}} \right) = 1/2 \rho_{0}$ to obtain the core radius as Eq. \ref{rc},
\begin{eqnarray} \label{rc}
	r_{\mathrm{c}} = a \cdot \left( 2^{2/\gamma} - 1 \right)^{1/2} \,,
\end{eqnarray}
same as Eq .22 in \cite{Elson87}. Radial density profile of IC 4756 and the analytical model with fitted parameters are presented in bottom panel of Fig. \ref{fig:rt_and_rdp}. The derived density $\rho_{0} = 0.325 \pm 0.130$ $\mathrm{M_{\odot}} \cdot \mathrm{pc}^{-3}$ is consistent with the central density from 644 members ($0.364$ $\mathrm{M_{\odot}} \cdot \mathrm{pc}^{-3}$), which is the density within 2 pc from the center. The core radius $r_{\mathrm{c}} =  4.33 \pm 0.75$ pc. The green belt represents $r_{\mathrm{c}}$ and its error. The curve matches up perfectly with the real density profile.

\section{Summary} \label{summary}

We discover the extended tidal tails of IC 4756 using the astrometry from {\it Gaia} EDR3. The method to determine the co-moving member candidates of IC 4756 contains defined neighbors in $\left( x, y, z, \kappa \cdot \mu_{\alpha}^{*}/\varpi, \kappa \cdot \mu_{\delta}/\varpi \right)$ and \texttt{HDBSCAN}, and other cuts in various parameter spaces. 644 member stars are identified as members of IC 4756. After distance corrections using a Bayesian approach, evident two tidal tails extend up to 180 pc in total length and present a clear S-shape in $\left( X^{\prime},Y^{\prime}\right)$ space. Hundreds of stars are found beyond the tidal radius, $ r_{\mathrm{t}} \sim 12.13$ pc. The shape and position of tidal tails are coincides with literatures. A clean sequence in CAMD and 2-D KS test further prove the validity of our selections of cluster members and the right age $\log \mathrm{t} = 8.987$ in literatures. Using mean mass in different annuli and the mass functions, we detect the evidence of mass segregation in this cluster. The radial mass density profile fits perfectly with EFF model, obtaining a core radius $r_{\mathrm{c}} = 4.33 \pm 0.75$ pc.

\acknowledgments

Xianhao Ye thanks Zhenyu Wu and Xiaoying Pang for helpful discussions. This study is supported by CMS-CSST-2021-B05, National Key R\&D Program of China No. 2019YFA0405502 and the National Natural Science Foundation of China under grant No. 11988101, 11973048, 11927804, 11890694. This
work has made use of data from the European Space Agency (ESA) mission {\it Gaia} (\url{https://www.cosmos.esa.int/gaia}), processed by the {\it Gaia} Data Processing and Analysis Consortium (DPAC, \url{https://www.cosmos.esa.int/web/gaia/dpac/consortium}). The Gaia archive website is \url{https://archives.esac.esa.int/gaia}. We also acknowledge the support from the 2m Chinese Space Station Telescope project.

\vspace{45mm}
%\facilities{}
\software{\texttt{Astropy} \citep{astropy13},
          \texttt{Galpy} \citep{Bovy15},
          \texttt{HDBSCAN} \citep{Campello13,McInnes17},
          \texttt{Matplotlib} \citep{matplotlib07},
          \texttt{Numpy} \citep{numpy11},
          \texttt{Pandas} \citep{pandas10},
          \texttt{Scipy} \citep{scipy20},
          \texttt{Scikit-learn} \citep{sklearn12},
          \texttt{Topcat} \citep{topcat05}
          }

% \appendix

% \section{Appendix information}

%\bibliography{sample63}{}
\bibliographystyle{aasjournal}

\end{document}